\documentclass[letterpaper]{IEEEtran} 
\usepackage{graphicx}
\usepackage{amsmath}
\usepackage{amssymb}
\usepackage{bm}
\usepackage{url}
\usepackage{array}

\begin{document}

\title{LoRa Throughput Analysis with Imperfect Spreading Factor Orthogonality}
\author{\authorblockN{Antoine Waret$^{*}$, Megumi Kaneko$^{*}$, Alexandre Guitton$^{\dagger}$, and Nancy El Rachkidy$^{\dagger}$\\
\authorblockA{$^{*}$ National Institute of Informatics, 2-1-2 Hitotsubashi, Chiyoda-ku, 101-8430 Tokyo, Japan \\
$^{\dagger}$ University Clermont Auvergne, CNRS, LIMOS, F-63000 Clermont-Ferrand, France} \\
Email: antoine.waret@grenoble-inp.org, megkaneko@nii.ac.jp, \{alexandre.guitton,nancy.el\_rachkidy\}@uca.fr} }

\maketitle

\vspace*{-0.5cm}
\begin{abstract}
LoRa is one of the promising techniques for enabling Low Power Wide Area Networks (LPWANs) for future Internet-of-Things (IoT) devices. Although LoRa allows flexible adaptations of coverage and data rates, it is subject to intrinsic types of interferences: co-SF interferences where end-devices with the same Spreading Factors (SFs) are subject to collisions, and inter-SF interferences where end-devices with different SFs experience collisions. Most current works have considered perfect orthogonality among different SFs. 
In this work, we provide a theoretical analysis of the achievable LoRa throughput in uplink, where the capture conditions specific to LoRa are included. Results show the accuracy of our analysis despite approximations, and the throughput losses from imperfect SF orthogonality, under different SF allocations. Our analysis will enable the design of specific SF allocation mechanisms, in view of further throughput enhancements.
\footnote{This work was supported by the Grant-in-Aid for Scientific Research (Kakenhi) no. 17K06453 from the Ministry of Education, Science, Sports, and Culture of Japan and by the NII Research Grant.}
\end{abstract}

\vspace*{-0.2cm}


\vspace*{-0.2cm}
\section{Introduction}

As the amount of mobile data traffic will rapidly increase during the upcoming years (studies forecast 50 billion Internet of Things (IoT) devices by 2020), new spectrum access strategies adapted to high device densities are ever more crucial. 
LoRa \cite{Sem15may} is one of the prominent candidates for Low Power Wide Area Networks (LPWANs), providing wide communication coverage with low power consumption, at the expense of data rate. Operating in license-free ISM bands (i.e., 868MHz in Europe), the LoRa PHY layer uses a chirp spread-spectrum modulation where different Spreading Factors (SFs) tune the chirp modulation rates. Lower SFs such as SF7 allow for higher data rates but reduced transmission range, whereas higher SFs such as SF12 provide longer range at lower data rates. On top of the LoRa PHY layer, the higher layers were defined by the LoRa Alliance and referred as LoRaWAN \cite{LoRaAll}. In particular, the MAC protocol is based on a pure ALOHA access with duty cycle limitations. The LoRaWAN network architecture is a star-like topology where end-devices communicate with gateways over several channels. 



Most studies on LoRa scalability so far assumed a perfect orthogonality among SFs, thereby creating virtual channels where multiple users with different SFs could simultaneously operate in the same channel and hence boost the achievable system throughput. 
Thus, a number of works have considered the effect of co-SF interference only, where end-devices using the same SF on the same channel are subject to collisions~\cite{Geo17apr}\cite{Bor16nov}. In particular, the outage probability of a LoRa system under co-SF interference was analyzed in~\cite{Geo17apr}, where a signal could be captured if its Signal-to-Interference-plus-Noise Ratio (SINR) was higher than 6 dB. As the number of devices increased, it was shown that those co-SF interferences were causing a scalability limit. 
However, recent studies have pointed out the fact that SFs were not perfectly orthogonal among themselves~\cite{Cro17}. 
Thus, the effect of inter-SF collisions was investigated through computer simulations and/or experiments.
Namely, \cite{Cro17}\cite{Zhu17} showed that inter-SF interferences could considerably decrease LoRa performance, especially for high SFs where frames have a greater time on air. 


In this work, we propose a theoretical analysis of the achievable throughput on the uplink of a LoRa network, encompassing the effects of co- and inter-SF interferences.  
To ensure a successful transmission, a packet must thus satisfy three conditions: 1) its SNR is above the reception threshold, 2) its SINR under co-SF interference is above the co-SF capture threshold, and its SINR under inter-SF interference is above the inter-SF capture threshold. 
Considering two different types of SF allocations, we theoretically derive the achievable throughput expressions for both perfect and imperfect SF orthogonality. Simulation results show the accuracy of our analytical expressions despite the necessary approximations, as well as the impact of the various types of interferences and SF allocations on the overall system performance.

 

\vspace*{-0.2cm}

\section{System Model}
\label{sec SysMod}

\begin{figure}[t]
\centering
\includegraphics[scale=0.27]{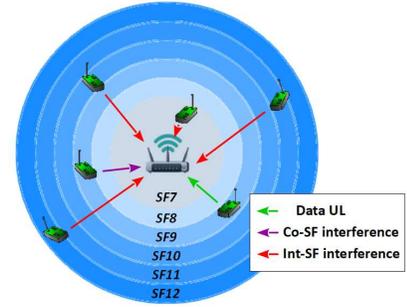}  
\vspace*{-0.4cm}
\caption{LoRa system setup - Case of \textit{SF-distance} allocation} 
\label{fig:system}
\vspace*{-0.7cm}
\end{figure}

We consider one cell of radius $R$ with one gateway located at its center, as depicted in Fig. \ref{fig:system}. There are \textit{N} end-devices uniformly distributed within the cell. We denote by $d_{i}$ the distance from end-device \textit{i} to the gateway. Since the goal of our analysis is to derive the achievable rate by LoRa, we assume that all end-devices transmit in a single channel of bandwidth $BW=125$kHz and that they all have packets to transmit. This corresponds to the pure ALOHA access as in LoRaWAN with saturated traffic\footnote{Our analysis can be easily applied to multiple channels and duty cycles.}. We consider $M=6$ SFs, for $m=\{m_{\mathrm{min}}, ...,m_{\mathrm{max}}\}$, with $m_{\mathrm{min}}=7$ and $m_{\mathrm{max}}=12$, with symbol times $T_{m}=\frac{2^{m}}{BW}$. The bit-rate $R_{m}$ of SF$_m$ is~\cite{Sem15may}  
\begin{equation}
R_{m}=\frac{{m} \times \mathrm{CR}}{\frac{2^{m}}{BW}}\nonumber,
\end{equation}
where CR is  the coding rate defined as $4/(4+n)$ with $n \in \{1,2,3,4\}$. Lower SFs allow higher data rate but lower communication range whereas higher SFs provide longer range at the expense of data rate (see Table \ref{table:lora_char}).

Two types of SF allocation will be investigated. In the first one, the spreading factors are uniformly distributed, i.e., every end-device has a probability $p_{m}=\frac{1}{M}$ of selecting SF$_m$. We refer to this allocation as \textit{SF-random}. In the second type of allocation referred to as \textit{SF-distance}, spreading factors are assigned according to the distance $d_{i}$. A device located inside the annulus defined by the smaller and larger circle radii $l_{m-1}$ and $l_{m}$, respectively, has SF$_{m}$. The distance threshold $l_{m}$ for SF$_m$ is given by $l_{m}=\left(\frac{P_0A(f_{c})}{\theta_{rx_{m}}}\right)^{\frac{1}{\alpha}}, $
where $A(f_{c}) = (f_{c}^{2} \times 10^{-2.8})^{-1} $ is the deterministic loss in the path loss model $L_{i}= \frac{A(f_{c})}{d_{i}^{\alpha}} $ as in \cite{ITUR}, $f_{c}$ the carrier frequency and $\alpha$ the path loss exponent. 
$\theta_{rx_{m}}$ is the receiver sensitivity of SF$_m$ (see Table \ref{table:lora_char}). All nodes transmit at equal power $P_{0}$. 
We assigned to $l_{6}$ and $l_{12}$ the origin of the cell and its radius respectively, i.e., $l_{6}=0$ and $l_{12}=R$. The ranges for each SF are given in Table \ref{table:lora_char}.
The probability of selecting SF$_m$ for the \textit{SF-distance} allocation is then given by $p_{m} = \int_{l_{m-1}}^{l_{m}} h(r)dr$,
where $h(r)$ is the pdf of the position of an end-device in the cell at distance $r$ from the gateway. For uniform distribution of devices within a cell of radius $R$, we get $h(r)=\frac{2r}{R^{2}}$.

The instantaneous SNR $\gamma_{i}$ of end-device $i$ is defined as $\gamma_{i} = P_{0}|h_{i}|^{2}L_{i}/\sigma_{n}^{2}$, where $|h_{i}|^{2}$ is the channel gain between end-device $i$ and the gateway (for Rayleigh fading, $h_{i}\sim\mathcal{CN}(0,1)$). $\sigma_{n}^{2} = -174+\mathrm{NF}+10\log(BW)$ [dBm] is the AWGN power and NF, the receiver noise figure.

Based on~\cite{Cro17}, it is assumed that in the event of a collision between frames of different SFs, one signal is received successfully if its SINR is higher than its ``InterSF capture threshold" in Table \ref{table:lora_char}. Moreover, if there are several signals with equal SFs transmitting on the same frequency simultaneously, the gateway is able to successfully receive one of them if its SINR is higher than 6 dB, for any SF$_m$~\cite{Geo17apr}\cite{Gou15}. Therefore, both types of interferences will be considered. 

\begin{table}
\begin{tabular}{|c|m{0.5cm}|m{1cm}|m{1cm}|m{1.2cm}|m{2cm}|}
\hline
\textbf{SF$_m$} & \textbf{Bit-rate} $R_{m}$ [kb/s] & \textbf{Receiver Sensitivity}\cite{Sem15may} $\theta_{rx_{m}}$ [dBm] & \textbf{Reception thresh.} $\mathrm{q_{SF}}_{m}$ [dB] & \textbf{InterSF capture thresh.}\cite{Cro17} $\mathrm{q_{iSF}}_{m}$ [dB] & \textbf{Dist. thresh.} $l_{m}$ [km]\\
\hline
7 & 5.47 & -123\centering & -6\centering & -7.5\centering & $0\sim0.453$  \\ \hline 
  8 & 3.13 & -126\centering & -9\centering & -9\centering & $0.453\sim0.538$  \\ \hline 
  9 & 1.76 & -129\centering & -12\centering & -13.5\centering & $0.538\sim0.639$ \\ \hline 
 10 & 0.98 & -132\centering & -15\centering & -15\centering & $0.639\sim0.760$  \\ \hline 
 11 & 0.54 & -134.5\centering & -17.5\centering &  -18\centering & $0.760\sim0.877$ \\ \hline 
 12 & 0.29 & -137\centering & -20\centering & -22.5\centering & $0.877\sim1$  \\ 
  \hline
\end{tabular}
\vspace*{-0.15cm}
\caption{LoRa Characteristics at $BW$ = 125 kHz, $\alpha=4$ and $R=1$ km} 
\label{table:lora_char}
\vspace*{-1.0cm}
\end{table}

\vspace*{-0.3cm}
\section{Proposed Throughput Analysis}
\label{sec Through}

An end-device's packet transmission in uplink is successfully received at the gateway if the three following conditions are fulfilled:

\textit{1) Reception condition:} 

Signal power must be above the SF-specific threshold $\mathrm{q_{SF}}_{m}$,
\begin{equation} 
P^{i}_{\mathrm{cap_{rx}}_{m}} = P(\gamma_{i} \geq \mathrm{q_{SF}}_{m}),
\label{eq:cap_rx}
\end{equation}
which is the probability that a received signal from end-device $i$ at a distance $d_{i}$ from the gateway has a SNR $\gamma_{i}$ above the threshold $\mathrm{q_{SF}}_{m}$ (Table I).

\textit{2) Co-SF capture condition:} 

The co-SF SINR of end-device $i$ is defined as
\begin{equation}
\mathrm{SINR}^{i}_{\mathrm{{coSF}}} = \frac{\gamma_{i}}{\sum_{k=1}^{K}\gamma_{k} + 1},
\end{equation}
where $K$ is the number of end-devices with the same SF as end-device $i$. A signal is successfully received if its $\mathrm{SINR}^{i}_{\mathrm{{coSF}}}$ is above $\mathrm{q_{coSF}}$. Therefore, the second condition is given by
\begin{equation}
P^{i}_{\mathrm{cap_{coSF}}} = P\left(\frac{\gamma_{i}}{ \sum{_{k=1}^{K} \gamma_{k}} + 1} \geq \mathrm{q_{coSF}}\right).
\label{eq:cap_co}
\end{equation} 

\textit{3) Inter-SF capture condition:}

As shown in~\cite{Cro17}\cite{Zhu17}, SFs are not perfectly orthogonal: a signal at SF$_m$ faces interferences from all $N-K$ signals on other SFs. The inter-SF SINR of end-device $i$ is defined as
\begin{equation}
\mathrm{SINR}^{i}_{\mathrm{{intSF}}} = \frac{\gamma_{i}}{\sum_{p=K+1}^{N}\gamma_{p} + 1}.
\end{equation}
A transmission is successful if $\mathrm{SINR}^{i}_{\mathrm{{intSF}}}$ is higher than the threshold $\mathrm{q_{iSF}}_{m}$, thus the third condition is given as 
\begin{equation}
P^{i}_{\mathrm{cap_{intSF}}_{m}} = P\left(\frac{\gamma_{i}}{\sum_{p=K+1}^{N}\gamma_{p} + 1} \geq \mathrm{q_{iSF}}_{m}\right).
\end{equation} 
\label{eq:cap_int}

Therefore, the uplink throughput $\tau$ can be expressed as
\begin{equation}
\tau = \sum_{m=m_{\mathrm{min}}}^{m_{\mathrm{max}}} R_{m} \, \times \, P_{\mathrm{success}}(\mathrm{SF}_{m}),
\label{eq:throughput}
\end{equation}
where $R_{m}$ is the bit-rate of SF$_m$ and $P_{\mathrm{success}}(\mathrm{SF}_{m})$, the probability of a successful transmission. 

Next, we analyze the throughput under perfect and imperfect SF-orthogonality, for the SF-distance case.

\subsection{Perfect Orthogonality}
We assume first that SFs are perfectly orthogonal, i.e., no end-device suffers inter-SF interferences. Hence, the probability of a successful transmission $P_{\mathrm{success}}(\mathrm{SF}_{m})$ is given by
\begin{equation}
P_{\mathrm{success}}(\mathrm{SF}_{m}) = \sum_{j=1}^{N} \binom{N}{j} p_{m}^{j}(1-p_{m})^{N-j} P(\mathrm{cap_{rx},cap_{coSF}}),
\label{eq:gen_perf}
\end{equation}
where $j$ denotes the total number of end-devices at SF$_m$, $\binom{N}{j} p_{m}^{j}(1-p_{m})^{N-j}$ is the probability of having $j$ end-devices among $N$ at SF$_m$ and $P(\mathrm{cap_{rx},cap_{coSF}})$ is the joint probability 
for reception condition and co-SF capture.

\subsubsection{For $j=1$}
the end-device is not subject to co-SF interferences, thus only the \textit{reception} condition needs to be satisfied,
\begin{equation}
P(\mathrm{cap_{rx},cap_{coSF}}) = P^{i}_{\mathrm{cap_{rx}}_{m}}.\nonumber
\end{equation}
First, we determine $P^{i}_{\mathrm{cap_{rx}}_{m}}$ for the SF-distance case. Given our assumptions, the SNR $\gamma_{i}$ is modeled as an exponential random variable with mean $\overline{\gamma_{i}}$. Therefore,

\begin{equation}
P^{i}_{\mathrm{cap_{rx}}_{m}}=P(\gamma_{i} \geq \mathrm{q_{SF}}_{m} | \overline{\gamma_{i}} ) \times P(\overline{\gamma_{i}})\nonumber,
\end{equation}
where $\mathrm{q_{SF}}_{m}$ is the specific threshold of SF$_m$. Defining $\overline{\gamma_{i}}=\frac{c}{r_{i}^{\alpha}}$, where $c=\frac{P_{0}\times A(f_{c})}{\sigma_{n}^{2}}$ is the path-loss constant, we can now rewrite,
\begin{equation}
P^{i}_{\mathrm{cap_{rx}}_{m}}= \int_{l_{m-1}}^{l_{m}} \exp\left(-\frac{\mathrm{q_{SF}}_{m}r_{i}^{\alpha}}{c}\right)\times \frac{2r_{i}}{R^{2}}dr_{i}.
\label{eq:res_rx}
\end{equation}
Although this integral cannot be expressed in closed form, it can be efficiently determined by numerical methods.

\subsubsection{For $j\geq2$} 
both the \textit{reception} and \textit{co-SF} conditions must be fulfilled. As $\mathrm{q_{SF}}_{m}$$\leq$1 in linear for all SFs whereas $\mathrm{q_{coSF}}_{m}=4$ (6 dB) for all SFs as explained in Section \ref{sec SysMod}, if co-SF capture is satisfied, so is the reception condition, hence
\begin{center}
\begin{tabular}{ccc}
	$P(\mathrm{cap_{rx},cap_{coSF}})$&=&$P^{i}_{\mathrm{cap_{coSF}}}$. \\
\end{tabular}
\end{center}
In case of co-SF interferences, there are $j$-1 interferers,
$P^{i}_{\mathrm{cap_{coSF}}} = P\left(\frac{\gamma_{i}}{ \sum{_{k=1}^{j-1} \gamma_{k}} + 1} \geq \mathrm{q_{coSF}}\right)$, which is developed using random instantaneous SNR variables $\gamma_{k}$ and random average SNR (position) variables $\overline{\gamma_{k}}$ as
\begin{center}
$\begin{array}{l}
P\left(\frac{\gamma_{i}}{ \sum{_{k=1}^{j-1} \gamma_{k}} + 1} \geq  \mathrm{q_{coSF}} | \overline{\gamma_{1}},...,\overline{\gamma_{i}},...,\overline{\gamma_{j-1}},\gamma_{1},...,\gamma_{j-1}\right)\\
\times P(\overline{\gamma_{1}},...,\overline{\gamma_{i}},...,\overline{\gamma_{j-1}},\gamma_{1},...,\gamma_{i-1},\gamma_{i+1},...,\gamma_{j-1}). \\
\end{array}$
\end{center}

\vspace*{0.3cm}
Marginalizing over $\gamma_{1},...,\gamma_{j-1}$ and making the change of variable $\overline{\gamma_{i}}=\frac{c}{r_{i}^{\alpha}}$, we get by independency of user channels,
\begin{eqnarray}
P^{i}_{\mathrm{cap_{coSF}}}  = \int_{l_{m-1}}^{l_{m}}\exp(-\frac{\mathrm{q_{coSF}} r_{i}^{\alpha}}{c}) \times \nonumber\\ 
\prod\limits_{\substack{p=1 \\ p\neq i}}^{j-1}\int_{l_{m-1}}^{l_{m}} \frac{h(r_{p})dr_{p}}{1+\mathrm{q_{coSF}}\left(\frac{r_{i}}{r_{p}}\right)^{\alpha}}  h(r_{i})dr_{i}\nonumber.
\end{eqnarray}

We define $I(r_{i}) = \int_{l_{m-1}}^{l_{m}} \frac{h(r)dr}{1+\mathrm{q_{coSF}}\left(\frac{r_{i}}{r_{p}}\right)^{\alpha}}$. Nodes are uniformly distributed within the cell, therefore $\prod_{p=1}^{j-1} I(r_{i}) = [I(r_{i})]^{j-1}$. As a result, the expression becomes,
\begin{equation}
P^{i}_{\mathrm{cap_{coSF}}} = \int_{l_{m-1}}^{l_{m}}\exp\left(-\frac{\mathrm{q_{coSF}}r_{i}^{\alpha}}{c}\right) [I(r_{i})]^{j-1}  h(r_{i})dr_{i}.
\label{eq:res_coSF}
\end{equation}

In particular, for $\alpha=4$, the primitive function of $I(r_{i})$ is
\begin{equation}
J(r_{i},r) = \left(\frac{r}{R}\right)^{2} - \left(\frac{r_{i}}{R}\right)^{2}\sqrt[]{\mathrm{q_{coSF}}}\arctan\left(\frac{1}{\left(\frac{r_{i}}{r}\right)^{2}\sqrt[]{\mathrm{q_{coSF}}}}\right).
\label{eq:prim_a4}
\end{equation}

In case of co-SF interferences, the interferers are the end-devices with the same SF as end-device $i$, thus with the same distance boundaries. The expression of $I(r_{i})$ becomes
\begin{equation}
I(r_{i})   =   J(r_{i},l_{m}) - J(r_{i},l_{m-1}).
\end{equation}

Therefore, (\ref{eq:gen_perf}) can be written for SF-distance allocation with perfect SF orthogonality as,
\begin{eqnarray}
P_{\mathrm{success}}(\mathrm{SF}_{m}) =  \binom{N}{1} (1-p_{m})^{N-1} \times P^{i}_{\mathrm{cap_{rx}}_{m}} \nonumber \\
+ \sum_{j=2}^{N} \binom{N}{j} (1-p_{m})^{N-j} \times P^{i}_{\mathrm{cap_{coSF}}},
\label{eq:res_perf}
\end{eqnarray}
with $P^{i}_{\mathrm{cap_{rx}}_{m}}$ given in (\ref{eq:res_rx}) and $P^{i}_{\mathrm{cap_{coSF}}}$ in (\ref{eq:res_coSF}).

\subsection{Imperfect Orthogonality}

In reality, spreading factors are not perfectly orthogonal, so all three capture conditions are to be satisfied to achieve a successful transmission. Thus, $P_{\mathrm{success}}(\mathrm{SF}_{m})$ becomes
\begin{eqnarray}
P_{\mathrm{success}}(\mathrm{SF}_{m}) = \sum_{j=1}^{N} \binom{N}{j} p_{m}^{j}(1-p_{m})^{N-j} \nonumber \\
\times P(\mathrm{cap_{rx},cap_{coSF},cap_{intSF}}),
\label{eq:gen_imp}
\end{eqnarray}
where $P(\mathrm{cap_{rx},cap_{coSF},cap_{intSF}})$ is the joint probability for reception condition, capture co-SF and capture inter-SF.

\subsubsection{For $j=1$} the end-device is only subject to inter-SF interferences and the \textit{reception} condition. As shown in Table \ref{table:lora_char}, the \textit{inter-SF} condition is dominant over the \textit{reception} one, especially as the number of end-devices increases. Thus,
\begin{equation}
P(\mathrm{cap_{rx},cap_{coSF},cap_{intSF}}) = P^{i}_{\mathrm{cap_{intSF}}_{m}}. \nonumber
\end{equation}
The expression of $P^{i}_{\mathrm{cap_{intSF}}_{m}}$ is similar to $P^{i}_{\mathrm{cap_{coSF}}}$, only with different thresholds and number of interferers. When dealing with inter-SF interferences, one must consider the end-devices within the cell that are not in the annulus corresponding to SF $m$, i.e., the end-devices with a different spreading factor. If $j$ is the number of end-devices at SF$_m$, then there are $N-j$ end-devices with other spreading factors. Therefore,
\begin{equation}
P^{i}_{\mathrm{cap_{intSF_{m}}}} = \int_{l_{m-1}}^{l_{m}}\exp\left(-\frac{\mathrm{q_{iSF_{m}}}r_{i}^{\alpha}}{c}\right) [\widetilde{I}(r_{i})]^{N-j} h(r_{i})dr_{i}.
\label{eq:res_intSF}
\end{equation}
Here, $\widetilde{I}(r_{i}) = \int_{\mathcal{R}\setminus\mathcal{R}_m}\frac{h(r)dr}{1+\mathrm{q_{coSF}}\left(\frac{r_{i}}{r_{p}}\right)^{\alpha}}$, where $\mathcal{R}\setminus\mathcal{R}_m$ denotes the whole cell area excluding the area corresponding to SF$_m$. Using (\ref{eq:prim_a4}), we obtain
\begin{equation}
\widetilde{I}(r_{i}) = J(r_{i},R)-J(r_{i},0)-\left[J(r_{i},l_{m}) - J(r_{i},l_{m-1})\right].
\label{eq:I_imp}
\end{equation}

\subsubsection{For $j\geq2$} all capture conditions are to be considered. As in the perfect orthogonality case, the \textit{reception} condition derives from the two others, thus
\begin{equation}
P(\mathrm{cap_{rx},cap_{coSF},cap_{intSF}}) = P(\mathrm{cap_{coSF},cap_{intSF}}). 
\label{eq:imp_j2}
\end{equation}
Because of the difficulty to find an exact expression of (\ref{eq:imp_j2}), as one condition might be dominant over the other depending on $j$, we approximate it as
\begin{equation}
P(\mathrm{cap_{coSF},cap_{intSF}}) \approx \mathrm{min}(P^{i}_{\mathrm{cap_{intSF}}_{m}},P^{i}_{\mathrm{cap_{coSF}}}) \nonumber.
\end{equation}

Finally, (\ref{eq:gen_imp}) can be written for SF-distance allocation with imperfect orthogonality as,
\begin{eqnarray}
&&P_{\mathrm{success}}(\mathrm{SF}_{m}) =  \binom{N}{1} \times P^{i}_{\mathrm{cap_{intSF}}_{m}} \nonumber \\
&&+\sum_{j=2}^{N} \binom{N}{j} (1-p_{m})^{N-j} \mathrm{min}(P^{i}_{\mathrm{cap_{intSF}}_{m}},P^{i}_{\mathrm{cap_{coSF}}}),\qquad
\label{eq:res_imp}
\end{eqnarray}
where $P^{i}_{\mathrm{cap_{intSF}}_{m}}$ is given in (\ref{eq:res_intSF}) and $P^{i}_{\mathrm{cap_{coSF}}}$ in (\ref{eq:res_coSF}).

Due to lack of space, the analytic throughput expressions for the SF-random case are not given, as they can be obtained similarly as for SF-distance. 

\vspace*{-0.3cm}
\section{Numerical Results}
\label{sec NumRes}

System throughput was evaluated for perfect/imperfect orthogonality as well as for both types of allocations, to assess the validity of our analytical expressions. The main parameters are $f_{c}=868$ MHz, $BW=125$ kHz and the end-devices' transmit power $P_{0}=14$ dBm. The path loss exponent was set to $\alpha=4$ as in~\cite{Geo17apr} (urban) and $R=1$ km.

\begin{figure}[t]
\centering
\includegraphics[scale=0.37]{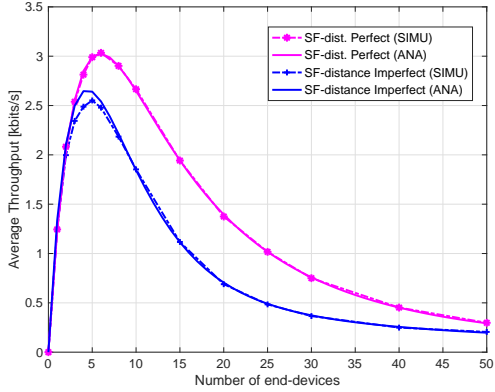} 
\vspace*{-0.5cm}
\caption{Throughput performance of SF-distance allocation for both perfect and imperfect orthogonality -- $R=1$km, $\alpha=4$}
\label{fig:sf_dist}
\end{figure}

Fig. \ref{fig:sf_dist} shows throughput performance obtained by simulations and by our theoretical derivations under SF-distance allocation for both perfect and imperfect orthogonality against varying numbers of end-devices transmitting simultaneously. We observe that our derived throughput expressions approach almost perfectly the simulations results with only a small interstice for small values of $N$ (around 5 end-devices) in the imperfect orthogonality case. Therefore, despite approximations, we obtain accurate throughput expressions for both orthogonality cases. Next, we can see that inter-SF interferences cause an early decrease of performance compared to the perfect orthogonality case. However, as the number of devices increases, co-SF interferences always lead to a scalability limit. These results show the impact of imperfect SF orthogonality over the system throughput, up to $50\%$ loss.

\begin{figure}[t]
\centering
\includegraphics[scale=0.35]{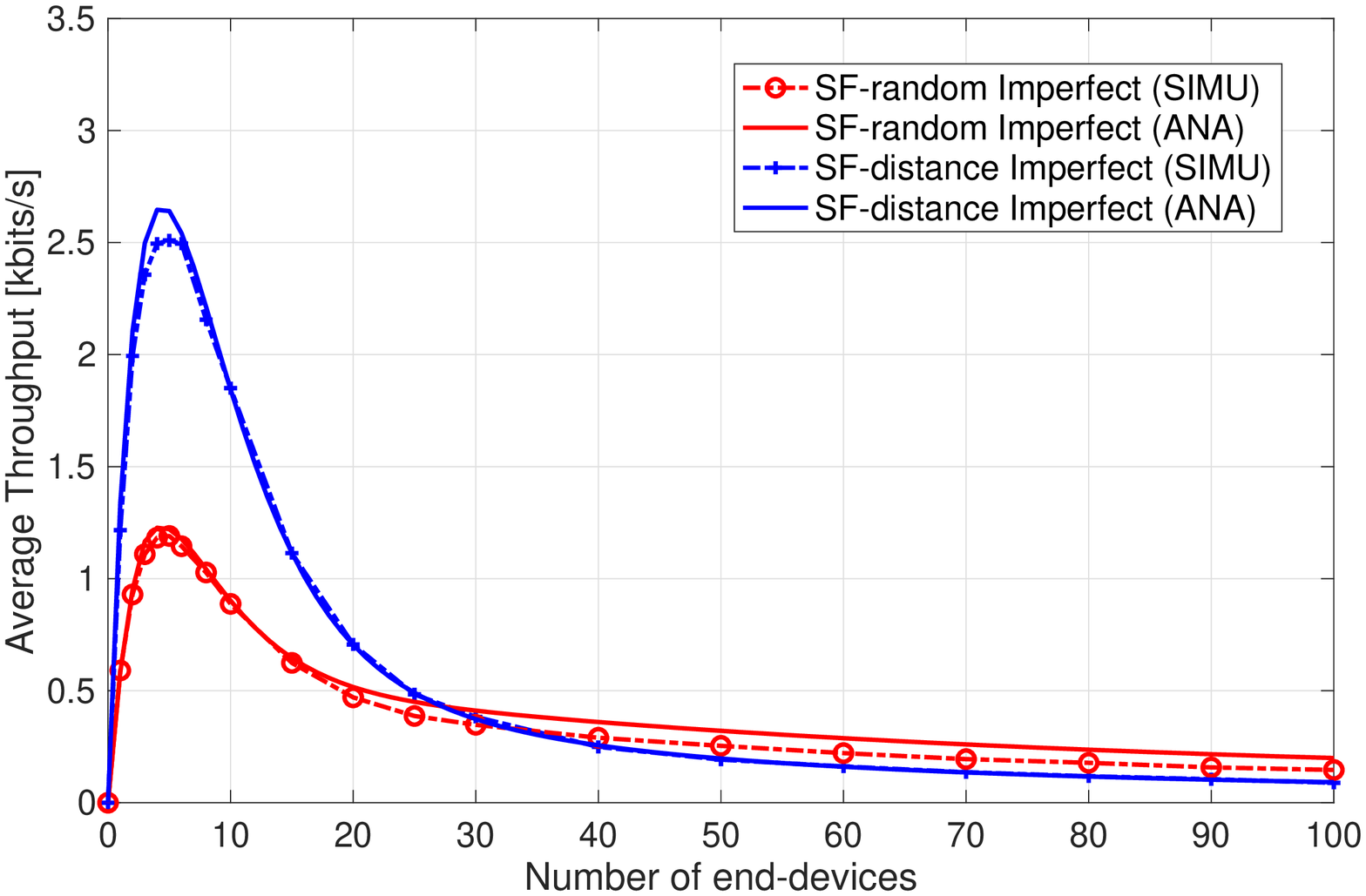}  
\vspace*{-0.85cm}
\caption{SF-random and SF-distance allocation throughput performances with imperfect orthogonality -- $R=1$km, $\alpha=4$}
\label{fig:uni_dist}
\end{figure}

Next, the throughput performance for both types of allocations with imperfect orthogonality are shown in Fig. \ref{fig:uni_dist}.
First, we can see that our analysis provides a good approximation of the achievable throughput under SF-random allocation. Then, we notice that for a small amount of end-devices (less than 25), better throughput efficiency is achieved in SF-distance allocation case (up to $100\%$ gain), since devices are more likely to satisfy the specific threshold $\mathrm{q_{SF}}_{m}$. On the other hand, SF-random allocation performs slightly better for greater values of $N$. Given the higher density of co-SF end-devices under SF-distance, these results suggest that even a simple SF-random policy provides a higher throughput as the number of end-devices increases. This is because the SF-random allocation favors the case where a limited number of devices randomly choose a small SF, by decreasing their collision probability. Moreover, small SFs lead to larger throughput than large SFs. Thus, our analysis will be useful to devise new allocation policies under various conditions and environments.

\vspace*{-0.3cm}
\section{Conclusion}
\label{sec conclusion}

We have considered the uplink of a single gateway LPWAN based on LoRa physical layer, for which a theoretical throughput expression was derived. Unlike most previous works, our analytical expression encompasses all three conditions required for successful frame transmission: SNR reception level, SINR level for co-SF capture, and SINR level for inter-SF capture. 
Results have shown the non-negligible impact of SFs' imperfect orthogonality, as well as the drastic effects of SF allocations on the overall throughput. Our analytic framework hence provides a precious tool for designing tailored SF allocations depending on environments and requirements, by predicting their impact on system performance.  




\vspace*{-0.3cm}
\addcontentsline{toc}{chapter}{References}
\bibliographystyle{IEEEtran}
\bibliography{IEEEabrv,LoRa_throughput_analysis}

\end{document}